\documentclass[10pt]{article}

\usepackage[utf8]{inputenc}
\usepackage[T1]{fontenc} 
\usepackage[english]{babel}
\usepackage{amsmath,amssymb,latexsym,eufrak,euscript}

\usepackage{algpseudocode}

\usepackage{url,tikz}
\usepackage{pgflibrarysnakes,pgfplots}

\usepackage{multicol}

\usepackage{authblk}

\usetikzlibrary{arrows}
\usetikzlibrary{automata}
\usetikzlibrary{snakes}
\usetikzlibrary{shapes}

\def \A {\mathcal{A}}

\def \P {\mathbb{P}}

\begin{document}
\label{firstpage}

  \title{An Approximation-based Approach for the Random Exploration of Large Models} 

\author[1]{Julien Bernard} 
\author[1]{Pierre-Cyrille Héam}
\author[1]{Olga Kouchnarenko} 
\affil[1]{ FEMTO-ST Institute, CNRS, Univ. Bourgogne Franche-Comt\'e,
  France, }

\maketitle

\begin{abstract}
System modeling is a classical approach to ensure their reliability since it is suitable both for a formal verification and for software
testing techniques. In the context of model-based testing an approach combining random testing and coverage
based testing has been recently introduced~\cite{DBLP:journals/sttt/DeniseGGLOP12}. 
However, this approach is not tractable on quite large
models. In this paper we show how to use statistical approximations to
make the approach work on larger models. Experimental results, on models of
communicating protocols, are provided; they are very promising, both for the
computation time and for the quality of the generated test suites.
\end{abstract}

\section{Introduction}\label{sec:intro}

Many critical tasks are now assigned to automatic systems. In this context,
producing trusted software is a challenging problem and a central issue in
software engineering. Recent decades have witnessed the strengthening of many formal approaches to ensure software reliability, from verification (model-checking, automatic theorem proving, static analysis) to testing, which remains an
inescapable step to ensure software quality. A great effort has been made by the scientific community in
order to upgrade hand-made testing techniques to scalable and proven framework.

Experience shows that random testing is a very efficient technique for
detecting bugs, especially at the first stages of testing activities. The strength of
random testing consists of its independence on tester's priority and choices. However,
the nature of random testing is {\it to draw randomly a test rather than
choosing it}, and it is therefore inefficient to detect behaviour of a program
occurring with a very low probability.
In~\cite{DBLP:journals/sttt/DeniseGGLOP12}, a random testing approach
consisting of the exploration of large graph based models has been proposed.
In order to tackle the problem of low probabilistic behaviour, the authors have also
suggested to bias the random generation, by combining it with a coverage
criterion, in order to optimize the probability to meet system' features described by this criterion.
It however requires the computation of large linear systems, which becomes rapidly intractable in practice for large
graphs.

In this paper we propose a sampling-based approach in order to compute
approximated values of the system' solutions, deeply improving the efficiency of the
computation. Experimental results on various graphs provided in the paper show a
very significant time computation improvement while keeping similar
covering statistical properties.

\subsection{Related Work}
A prevailing methods in model-based testing consists in designing the system
under test by a graph-based formal
model~\cite{DBLP:journals/stvr/Wah96,lee1996principles} on which different
algorithms may be used to generate the test suites. This approach has been
used for a large class of applications from security of Android
systems~\cite{DBLP:conf/icst/TangCZGXHBM17} to digital
ecosystems~\cite{DBLP:conf/icst/LimaF16}. A large variety of models can be
used for model-based testing such as Petri
nets~\cite{DBLP:conf/icst/ThummalaO16}, timed
automata~\cite{DBLP:conf/icst/WangPB17}, pushdown
automata~\cite{DBLP:journals/stvr/DreyfusHKM14}, process
algebra~\cite{DBLP:journals/infsof/AlbertoCGS17}, etc. Moreover, a strength
of model-based testing is that it can be combined with several verification
approaches, such as model-checking~\cite{DBLP:journals/stvr/DadeauHKMR15} or
those using SMT-solvers~\cite{DBLP:conf/tap/AichernigJK13}. A general
taxonomy with many references on model-based testing approach can be found
in~\cite{DBLP:journals/stvr/UttingPL12}.

Random testing approaches have been introduced
in~\cite{DBLP:conf/icse/DuranN81} and are widely used in the literature,
either for generating
data~\cite{DBLP:conf/pldi/GodefroidKS05,DBLP:conf/icst/HeamN11} or for
generating test suites~\cite{DBLP:conf/soqua/Oriat05}. As far as we know, the
first work combining random testing and model-based testing has been
proposed in~\cite{DBLP:conf/issta/GroceJ08} as a combination of
model-checking and testing. In~\cite{DBLP:journals/sttt/DeniseGGLOP12} the
authors have proposed an improved approach to explore the models at random.
This technique has been extended to pushdown
models~\cite{DBLP:conf/tap/HeamM11,DBLP:journals/stvr/DreyfusHKM14} and to
grammar-based systems~\cite{DBLP:conf/icst/DreyfusHK13}.

\subsection{Formal Background}

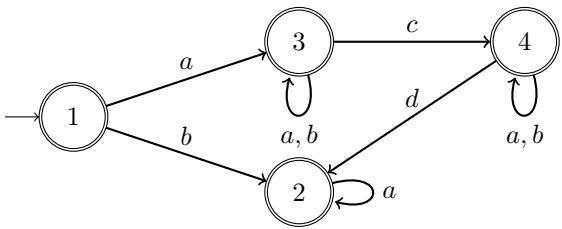
\begin{figure}[!b]
\begin{center}
\begin{tikzpicture}
\node[draw, state, accepting, initial, initial text=] (1) at (0,0) {$1$};
\node[draw, state, accepting] (3) at (3,1) {$3$};
\node[draw, state, accepting] (4) at (6,1) {$4$};
\node[draw, state, accepting] (2) at (3,-1) {$2$};

\path[->,thick] (1) edge[above] node {$b$}  (2);
\path[->,thick] (1) edge[above] node {$a$}  (3);
\path[->,thick] (3) edge[above] node {$c$}  (4);
\path[->,thick] (4) edge[above] node {$d$}  (2);

\path[->,thick] (2) edge[loop right] node {$a$}  ();
\path[->,thick] (3) edge[loop below] node {$a,b$}  ();
\path[->,thick] (4) edge[loop below] node {$a,b$}  ();

\end{tikzpicture}
\end{center}
\caption{Illustrating example.}\label{fig:toyexample}
\end{figure}

For a general reference on probability theory, see~\cite{DBLP:books/daglib/0012859}.

\paragraph{Finite Automata.}
Models considered in this article are finite automata, that are labelled
graphs. More precisely, a finite automaton $\A$ is a tuple
$(Q,\Sigma,E,I,F)$, where $Q$ is a finite set of states, $\Sigma$ is a
finite alphabet, $E\subseteq Q\times \Sigma\times Q$ is the set of
transitions, $I\subseteq Q$ is the set of initial states, and $F\subseteq Q$
is the set of final states. A path $\sigma$ in a finite automaton is a
sequence $(p_0,a_0,p_1)\ldots (p_{N-1},a_{N-1},p_N)$ of transitions. The integer
$N$ is the length of the path. If $p_0\in I$ and $p_N$ in $F$, $\sigma$ is
said successful. The path $\sigma$ visits a state $q$ if there exists $i$
such that $p_i=q$. An automaton is trim if every state is visited by at
least one successful path. All automata considered throughout this paper are
trim. An example of an automaton is depicted in Fig.~\ref{fig:toyexample}~: its
set of states is $\{1,2,3,4\}$, the alphabet is $\{a,b,c,d\}$, its set of
transitions is $$\{(1,a,3),(3,a,3), (3,b,3), (3,c,4), (4,a,4),
(4,b,4), (1,b,2), (2,a,2),\\ (4,d,2)\},$$ its set of initial states is reduced
to $\{1\}$ and all its states are final.

\begin{figure}[!t]
\begin{center}
\begin{tikzpicture}
\node[draw, state, initial, initial text=] (1) at (0,0) {$1,0$};
\node[draw, state, accepting] (3) at (3,1) {$3,0$};
\node[draw, state] (4) at (6,1) {$4,0$};
\node[draw, state] (2) at (3,-1) {$2,0$};

\path[->,thick] (1) edge[above] node {$b$}  (2);
\path[->,thick] (1) edge[above] node {$a$}  (3);
\path[->,thick] (4) edge[above] node {$d$}  (2);
\path[->,thick] (2) edge[loop right] node {$a$}  ();

\path[->,thick] (4) edge[loop below] node {$a,b$}  ();

\node[draw, state, accepting] (11) at (0,-4) {$1,1$};
\node[draw, state, accepting] (31) at (3,-3) {$3,1$};
\node[draw, state, accepting] (41) at (6,-3) {$4,1$};
\node[draw, state, accepting] (21) at (3,-5) {$2,1$};

\path[->,thick] (3) edge[ bend right, left, pos=0.7] node {$a,b$}  (31);
\path[->,thick] (3) edge[right] node {$c$}  (41);
\path[->,thick] (11) edge[above] node {$b$}  (21);
\path[->,thick] (11) edge[above] node {$a$}  (31);
\path[->,thick] (31) edge[above] node {$c$}  (41);
\path[->,thick] (41) edge[above] node {$d$}  (21);

\path[->,thick] (21) edge[loop right] node {$a$}  ();
\path[->,thick] (31) edge[loop below] node {$a,b$}  ();
\path[->,thick] (41) edge[loop below] node {$a,b$}  ();

\end{tikzpicture}
\end{center}
\caption{Illustrating example for constrained paths.}\label{fig:toy2}
\end{figure}
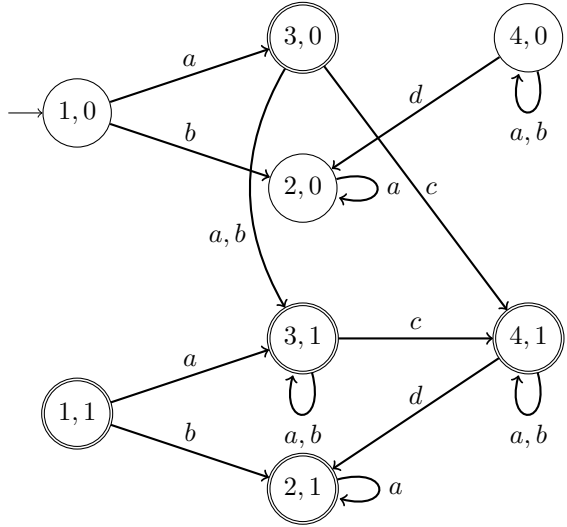

Let $\A=(Q,\Sigma,E,I,F)$ be a $n$-state automaton and $q\in Q$. We denote
by $\A_q$ the automaton on the alphabet $\Sigma$ whose set of states is
$(Q\times\{0,1\}$ (two copies of $Q$) and~:
\begin{itemize}
\item Its set of initial states is $I\times\{0\}$,
\item Its set of final states is $F\times\{1\}\cup
(F\cap\{q\})\times\{0\}$,
\item Its set of transitions is  $E^\prime=\{((p,0),a,(p^\prime,0))\mid
(p,a,p^\prime)\in E\text{ and }p\neq q\}\cup \{((p,1),a,(p^\prime,1))\mid
(p,a,p^\prime)\in E\}\cup \{((q,0),a,(p^\prime,1))\mid (q,a,p^\prime)\in
E\}$.
\end{itemize}
Intuitively, a successful path in $\A_q$ starts with an initial state of the
 form $(q_0,0)$ and remains in a state of the form $(p,0)$ until it visits
 $q$. Then, if $q$ is final in $\A$ it may ends or continue with states of
 the form $(p,1)$. One can easily show that there is a bijection between the
 set of successful paths of $\A_q$ of length $N$ and the set of successful
 paths of $\A$ of length $N$ visiting $q$. 
 denoted
Let us consider for instance the automaton depicted in Fig.~\ref{fig:toyexample} and
state~3. The corresponding automaton is depicted in
Fig.~\ref{fig:toy2} (states~(4,0), (1,1) and~(2,0) have to be removed to make the
resulting automaton trim).

Automata used in many testing applications  have a bounded outgoing
degree. Throughout this paper, we consider that $|E|=O(|Q|)$. Note that it is
not a theoretical requirement: it is only used for the complexity issues.
Indeed, all proposed algorithms work for any automata. Under this hypothesis,
computing $\A_q$ can be done in time $O(n^2)$ and the resulting automaton
has at most twice the number of states (regarless of the fact that $|E|=O(|Q|)$).

\paragraph{Counting Paths.}
We call {\sc NumPaths} an algorithm that, given a finite automaton $\A$ and
a positive integer $N$, computes the number of successful paths of
length $N$ in $\A$. We call {\sc RandomPath} an algorithm that, given a finite
automaton $\A$ and positive integers $N,k$, randomly, uniformly and
independently generates $k$ successful paths of length $N$ in $\A$. Several
algorithms have been developed for processing {\sc NumPaths} and {\sc
RandomPath}~\cite{DBLP:journals/tcs/OudinetDG13}, whose complexities depend
on several parameters. Let us observe, without going into details, using floating
point arithmetics, that {\sc NumPaths} can be performed in $O(nN\log N)$, where
$n$ is the number of states of $\A$. And {\sc RandomPath} can be performed
in time $O(knN\log^2N)$. Note that the different approaches may have
different meanings of time/space complexities, both for the preprocessing step and
the generation step. The reader can see~\cite[Table 4]{DBLP:journals/tcs/OudinetDG13}
and~\cite[Table~1]{DBLP:journals/algorithmica/BernardiG12} for more details.

\paragraph{Random Biased Exploration of Finite Automata.}
The objective is here to biased the random generation of paths (i.e. not use
a uniform random generation) in order to improve the state coverage of the
automata. It is necessary to provide a quite detailed description of the
algorithms in~\cite{DBLP:journals/tcs/OudinetDG13}. The first approach,
denoted later {\tt Uniform}, consists in uniformly picking up a given number
of paths from the set of successful paths of a given length. The approach
can be applied to very large graphs with hundreds of nodes
(see~\cite[Section~6]{DBLP:journals/sttt/DeniseGGLOP12}). However, rare
events can be missed up, and in order to optimize\footnote{Computing test
suites of a reduced size is a major issue in the testing process, since
executing test on the system is frequently a complex issue (not adressed in
this paper).} the coverage criterion (let us present it here for nodes
coverage\footnote{The approach can easily be adapted for transitions
coverage.}) of the graph, the following approach, denoted later {\tt Exact},
is proposed to produce $k$ successful paths of an automaton $\A$ whose set
of states is $\{1,\ldots,n\}$:

\begin{enumerate}
\item Choose a set $S$ of successful paths (for instance those of length less than or
equal to a constant $N$),
\item For each pair of nodes, compute the probability $\alpha_{i,j}$ that a
path of $S$ visiting $j$ also visits $i$,
\item Solve the linear programming system whose variables are
$p_{\rm min},\pi_1,\ldots,\pi_n$:
\begin{equation}\label{lps}
\begin{array}{l}
\text{maximize  }p_{\rm min},\text{ under the constraints}\\
\left\{
\begin{array}{l}
\text{for all } j,\ p_{\rm min}\leq \sum_{i=1}^n \alpha_{i,j}\pi_i\\
1=\sum_{i=1}^n\pi_i
\end{array}
\right.
\end{array}
\end{equation}

Solution is a distribution $\pi=(\pi_1,\ldots,\pi_n)$ of probabilities over the
states  of the automaton,
\item  Repeat $k$ times: pick a node $i$ up at random according to the
distribution $\pi$. Pick up at random (uniformly) a path visiting $i$.
\end{enumerate}

The goal of the linear programming system is to optimize the minimal
probability $p_{\rm min}$ of a state to be visited by a random path. 

Let us illustrate this approach on the example depicted in Fig.~\ref{fig:toyexample}.
Note that if the goal is to cover a given proportion of the set of states
(for instance) Step 4. can be replaced by: generate paths as soon as the
wanted proportion of states are visited by these paths.
There are $16$ successful paths of strictly positive length less than or equal
to~$3$ reported in Table~\ref{tab:examplepaths}. Since the automaton is
deterministic, one can identify successful paths with their labels. Let
$S_{\rm exa}$ be this set of paths.

\begin{table}[!t]
\begin{center}
\begin{tabular}{|c|l|c|}
\hline
length & paths & number of paths\\ \hline \hline
1 & $a$, $b$ & 2\\ \hline
2 & $aa$, $ab$, $ac$, $ba$& 4\\ \hline
3 & $aaa$, $aab$, $aba$, $abb$, $aac$, $abc$, $aca$, $acb$, $acd$, $baa$& 10\\ \hline
\end{tabular}
\end{center}
 \caption{Successful paths of length less than or equal to 3 for
   Example~\ref{fig:toyexample}.}\label{tab:examplepaths}
\end{table}

There are $4$ out of $16$ paths of $S_{\rm exa}$ visiting state~$2$.
Therefore, the probability of visiting state~$2$ by uniformly generated paths of $S_{\rm exa}$ is~$\frac{1}{4}$. In order to generate a path
visiting $2$, one has to generate averagely $4$ tests.
Moreover, for this example $\alpha_{i,j}$'s matrix is
$$
\left(
\begin{array}{cccc}
1 & 1 & 1 & 1\\
0.25 & 1 & \frac{1}{13} & \frac{1}{6} \\
0.825 & 0.25 & 1 & 1 \\
0.375 & 0.25 & \frac{6}{13} & 1 \\
\end{array}
\right).
$$
For instance, $\alpha_{1,i}=1$ for every $i$ since all paths visit $1$.
Similarly, $\alpha_{3,4}=1$ since all paths visiting $4$ also visits $3$.
There are four paths ($b$, $ba$, $baa$ and $acd$) of needed length visiting
$2$ and, among these paths, only $acd$ visits $4$. Therefore
$\alpha_{4,2}=\frac{1}{4}$. The resolution\footnote{Resolutions have been
performed using the {\tt lp\_solve} solver.} of linear programming
systems~(\ref{lps}) provides in this context the solution: $\pi_1=0$,
$\pi_2=0.526315$, $\pi_3=0$, and $\pi_4=0.473685$. In this context, the
biased approach covers all states averagely with less than $3$ generated
paths.

The bottleneck of this approach is Step~2. since computing the
$\alpha_{i,j}$s requires many manipulations on the graphs (it requires to
compute the ${(\A_i)}_j$): for each $i\neq
j$, Algorithm~{\sc NumPaths} has to be applied to graphs 4 times larger
than the initial ones. The complexity is in $O(n^3N\log N)$ with quite large
involved constants, making the approach intractable for big $n$'s.

\subsection{Contributions}

In this paper we propose to not exactly compute the $\alpha_{i,j}$s but
instead to approximate them by using statistical sampling, as described in
Section~\ref{sec:contrib}. Experimental results on several examples of
communication protocols models are provided in Section~\ref{sec:xp}. The
paper reports on very promising experimental results: the computation time
is significantly better for a similar quality of the large graphs coverage.

\section{Approximating the Linear Programming Systems}\label{sec:contrib}

In this section, we propose to approximate the coefficients $\alpha_{i,j}$
by $\alpha_{i,j}^{\rm approx}$ by using classical sampling techniques. Using
$m$ times Algorithm {\tt RandomPath}, one can count as $m_i$ the number
of paths visiting $i$, and $m_{i,j}$ the number of paths visiting both $i$
and $j$. If $m_i\neq 0$ then $\alpha_{i,j}^{\rm approx}=\frac{m_{i,j}}{m_j}$.

\subsection{Approximation Algorithm}
More precisely, let there be a trim finite automaton $\A=(Q,A,E,I,F)$, a
strictly positive integer $m$, a strictly positive integer $N$ and a
strictly positive integer~$r$ (the
parameter $r$ is used to provide some bounds on the precision of the
approximation: each evaluation of a parameter is estimated using a sample of
size at least~$r$).

\begin{itemize}
\item []{\bf (Step 1):} Generate  $m$ successful paths in $\A$ of length
  less than or equal to $N$ uniformly. For each $i\in Q$, let $m_{i}^{\rm approx}$ be the
  number of these paths visiting $i$, and  $m_{i,j}^{\rm approx}$ be the
  number of these paths visiting both $i$ and $j$.
\item[]{\bf (Step 2):} For each $i,j\in Q$, $i\neq j$
\begin{itemize}
\item[(a)] If $r=0$ and $m_j^{\rm approx}=0$, then  let $\alpha_{i,j}^{\rm approx}=0$,
\item[(b)] If $m_{j}^{\rm approx} > r$, let $\alpha_{i,j}^{\rm approx}=m_{i,j}^{\rm approx}/m_{j}^{\rm approx}$,
\item[(c)] If  $m_{j}^{\rm approx} \leq  r$, generate $r$ paths visiting $i$ and
  set $\alpha_{i,j}^{\rm approx}$ as the proportion of these paths visiting
  $j$.
\end{itemize}
\item[]{\bf (Step 2):} For each $i\in Q$, $\alpha_{i,i}^{\rm approx}=1$.
\end{itemize}

Let us illustrate the approach on the example depicted in
Fig.~\ref{fig:toyexample}, with $N=4$ and $r=0$. Rather than compute exactly
the $\alpha_{i,j}$'s, we randomly and uniformly generate 1000 paths of length
less than or equal to~3. We obtain the following matrix for the
$\alpha_{i,j}^{\rm approx}$~:
$$
\left(
\begin{array}{cccc}
1 & 1 & 1 & 1\\
0.243 & 1 & 0.0835 &  0.1778\\
0.826 & 0.284 & 1 & 1 \\
0.288 & 0.284 & 0.4697 & 1 \\
\end{array}
\right).
$$
The resolution of systems provides the solution $\pi_1=0$, $\pi_2=0.538019$,
$\pi_3=0$ and $\pi_4=0.461981$.

In this example, there are 243~paths visiting $2$, 826 paths visiting $3$
and 288 paths visiting $4$. Therefore, running the algorithm with $r=250$
will change  the second column of the matrix since $m_2^{\rm
approx} < 250$. In this case, the automaton for the paths visiting $2$ is
computed. Generating $250$ paths visiting state~$2$ provides the following
matrix:
$$
\left(
\begin{array}{cccc}
1 & 1 & 1 & 1\\
0.243 & 1 & 0.0835 &  0.1778\\
0.826 & 0.256 & 1 & 1 \\
0.288 & 0.256 & 0.4697 & 1 \\
\end{array}
\right).
$$
The resolution of systems provides the solution $\pi_1=0$, $\pi_2=0.524965$,
$\pi_3=0$ and $\pi_4=0.475035$.

Section~\ref{sec:xp} describes more experiments and provides details,
both on the quality of the results and on the time to compute the
$\alpha_{i,j}$'s.

Notice too that, as mentioned in~\cite{DBLP:journals/sttt/DeniseGGLOP12},
the optimal solution leads to a loss of randomness: many $\pi_i$'s a null.
It is proposed in~\cite{DBLP:journals/sttt/DeniseGGLOP12} to fix minimal
probability to the $\pi_i$'s. It can be directly adapted in our approach by
adding, in the programming linear system, some inequations of the form
$\pi_i\geq \varepsilon$. This situation would not be considered in the
experiments developed in this paper. 
\subsection{Complexity}

We investigate in this section the worst case complexity of the proposed algorithm.
Step~(1) can be performed in time  $O(mnN\log^2N+mn^2)$: first the $m$ paths are
generated in time  $O(mnN\log^2N)$. These paths are not stored but a table $t$
%
of size $m\times n$ is filled in the following way: $t[i][j]=1$
if the $i$-th path visits state~$j$, and $t[i][j]=0$ otherwise. It is done on
the fly and in time $O(nm)$. The $m_i^{\rm approx}$ are calculated by computing
columns sums in time $O(nm)$ too. Similarly, each  $m_{i,j}^{\rm approx}$
can be computed in time $O(m)$. Therefore, computing all of them is performed
in time $O(mn^2)$.

Step~2-(a) is performed in time $O(1)$ as well as Step~2-(b). Step~2-(c) is
performed in time $O(rn N \log^2 N)$: computing the specific automaton is
done in time $O(n)$ (under the hypothesis that the number of transitions is in
$O(n)$).

Step~3 is performed in time $O(n)$.

In conclusion, if we denote by $s$ the number of calls to Step~2-(c),
the complexity is: $O(((sr+m)nN\log^2N+mn^2)$.

A small $r$ (for instance $r=0$) will provide a small $s$ ($s=0$), but a
coarser approximation, as exposed in the next section.

\subsection{Precision of the $\alpha_{i,j}^{\rm approx}$'s}

Each $\alpha_{i,j}$ is the parameter of Bernoulli's Law (see~\cite[Section~2.2]{DBLP:books/daglib/0012859}). The precision of
the estimation can classically be obtained using either Bienaym\'e-Chebyshev's
Inequality~\cite{bienayme,cheb} or Hoeffding's Inequality~\cite{hoeff}.

First, assuming that $m_{j}^{\rm approx}> r$, then Bienaym\'e-Chebyshev's
Inequality provides for any $\varepsilon> 0$:

$$\P(|\alpha_{i,j}^{\rm
approx}-\alpha_{i,j}|\geq \varepsilon)\leq \frac{\alpha_{i,j}(1-\alpha_{i,j})}{\varepsilon^2{m_j^{\rm
approx}}}\leq \frac{1}{4\varepsilon^2r},$$
and if $m_{j}^{\rm approx}\leq  r$, it provides:
$$\P(|\alpha_{i,j}^{\rm
approx}-\alpha_{i,j}|\geq \varepsilon)\leq \frac{\alpha_{i,j}(1-\alpha_{i,j})}{\varepsilon^2r^2}\leq \frac{1}{4\varepsilon^2r}.$$
In order to have an $\varepsilon=0.1$ precision with a 0.95 confidence
level, $r$ has to be fixed to $500$ (this is an upper bound).

Secondly, one can have another evaluation using Hoeffding's Inequality
(better in most of cases): for any $0< \rho < 1$,

\begin{equation*}\label{eq-approx-1}
\mathbb{P}\left(|\alpha_{p,q}^{\rm approx}-\alpha_{p,q}|\geq \varepsilon \right)\leq
2e^{-2r\varepsilon^2}.
\end{equation*}

In order to have an $\varepsilon=0.1$ precision with a 0.95 confidence
level, $r$ has to be fixed to $185$ (this is also an upper bound).

Let us note that the two above inequalities provide upper bounds that
are not very tight: for states $j$ frequently visited by random paths,
$m_j$ will be significantly greater than $r$, and the estimation of
the algorithm will be very precise. As it is shown in the next
section, running the algorithm with $r=0$ frequently provides very
acceptable solutions and very good solutions with $r=10$. For $r=10$
the two bounds above do not ensure precise estimations: Hoeffding's
Inequality states that with a 0.8 confidence level we have an
estimation of $\alpha_{i,j}$ with $\varepsilon=0.34$. An hypothesis
explaining why $r=10$ works is that it is important to detect whether
while visiting $j$ the probability to also visit $i$ is
significant. But it's not critical to know how significtant it is, for
instance if $\alpha_{i,j}=0.1$ or $0.4$; it is important to know that
generating a path visiting $j$ will quite frequently provide a path
visiting~$i$.

Finally, other statistical tools can be used to obtain bounds on $r$, for
instance the well-known central limit theorem.

\section{Experiments}\label{sec:xp}

\begin{figure}
{\small
\begin{tabular}{|l||c|c|c|c|c|}
\hline
r=0 &  $90\%$ &   $95\%$ &   $99\%$ &   $100\%$ \\\hline\hline
{\tt RW}& 4.38 2--10&  4.38 2--10&  4.38 2--10&  4.38 2--10\\\hline
{\tt Uniform}& 4.22 2--11& 4.22 2--11& 4.22 2--11& 4.22 2--11\\\hline
{\tt Approx~10}& 4.65 2--13&  4.65 2--13& 4.65 2--13 &  4.65 2--13\\\hline
{\tt Approx 1000}& 4.18 2--12& 4.18 2--12& 4.18 2--12& 4.18 2--12\\\hline
{\tt Exact}& 4.42 2--12& 4.42 2--12& 4.42 2--12& 4.42 2--12\\\hline
\end{tabular}
\hspace{0.5cm}
\begin{tabular}{l}
{\sf Barber1}\\
15 states\\
18 transitions\\
\end{tabular}

\bigskip

\begin{tabular}{|l||c|c|c|c|c|}
\hline
r=0 &  $90\%$ &   $95\%$ &   $99\%$ &   $100\%$ \\\hline\hline
{\tt RW}& 428 18--1724& 790 159 -- 2062& 2451.4 599--7727&2451.4 599--727\\\hline
{\tt Uniform}  & 22.58 9--45& 35.45 14--79& 92.36 32--216&92.36 32--216\\\hline
{\tt Approx~10}& 11.7 7--21 &17.2 10--52 & 39.3 16-181&
39.35 16 -- 181\\\hline
{\tt Approx 1000}& 10.46 7--20& 14.9 8.31& 30.9 11--73& 30.9 11--73\\\hline
{\tt Exact} & 11.23 7--24& 15.5 8--30& 29.8 13--71& 29.8 13--71\\\hline
\end{tabular}

{\sf Dekker1}, 86 states, 178 transitions
\bigskip

\begin{tabular}{|l||c|c|c|c|c|}
\hline
 r=0 &  $90\%$ &   $95\%$ &   $99\%$ &   $100\%$ \\\hline\hline
{\tt RW}& 15 9-27& 21.5 13--37&37.8 18--70& 50.5 19--130\\\hline
{\tt Uniform}   & 31.2 16--51& 47.7 22--88& 87.6 40--166& 115.5 45--278\\\hline
{\tt Approx~10}& 19.5 13--34& 28.1 17-51& 48.6 27--146& 67.9 30--146\\\hline
{\tt Approx 1000}& 18 10--29& 24.5 13-38& 40.3 20--74& 50.0 26-86 \\\hline
{\tt Exact} &18.4 12--30 & 24.8 18--38& 39.5 23--63& 49.6 26--102\\\hline
\end{tabular}
\begin{tabular}{l}
{\sf Fsm1}, 120 states, 582 transitions
\end{tabular}
\bigskip

\begin{tabular}{|l||c|c|c|c|c|}
\hline
r=0 &  $90\%$ &   $95\%$ &   $99\%$ &   $100\%$ \\\hline\hline
{\tt RW}& 102.1 8-430& 178.3 16--735& 330.6 16--1125& 330.6 16--1125\\\hline
{\tt Uniform} & 9.7 2--35& 13.1 2--39& 21.9 2--90&21.9 2--90\\\hline
{\tt Approx~10}& 2.9 2--11& 3.5 2--11& 5.1 2--23& 5.1 2--23\\\hline
{\tt Approx 1000}& 3.0 2--8& 3.0 2--8& 3.5 2--8&3.5 2--8 \\\hline
{\tt Exact} & 3.3 2--9& 3.3 2--9& 3.7 2--9& 3.7 2--9\\\hline

\end{tabular}
\smallskip

{\sf Moesi2},  22 states, 43 transitions

\bigskip
\begin{tabular}{|l||c|c|c|c|c|}
\hline
r=0 & $50\%$ &  $90\%$ &   $95\%$ &   $99\%$ &   $100\%$ \\\hline\hline
{\tt RW}& 11.5 6--18& 80.6 45--153& 122.8 73--247& 260.8
120--514&342.4 156--649\\\hline
{\tt Uniform}   & 8.78 7--13& 54.4 36--82& 80.9 51--12& 180.972--333& 277.3 105--776\\\hline
{\tt Approx~10}& 7.49 5--10& 37.6 27--56& 53.3 38-91& 102.6 56--205&
140.6 59--347\\\hline
{\tt Approx 1000}& 7.6 6 -- 11&35.8 24--51 &50.9 32--80 & 88.11 58-147&
106.8 62--179 \\\hline
{\tt Exact} & 7.6 5--10& 34.9 24--46& 47.6 33--63& 82.3 55--121& 101.5 56--165\\\hline

\end{tabular}
\smallskip
{\sf Kanban1},  160 states, 1151 transitions

}
\caption{Comparative results (1) for number of generated tests}\label{fig:res1}
\end{figure}

\begin{figure}[!h]
{\small

\begin{tabular}{|l||c|c|c|}
\hline
r=0 &  $90\%$ &   $95\%$   \\\hline\hline
{\tt RW} & 54991.6 39414--69917& 155803 117044-214680\\\hline
{\tt Uniform}  & 937.1 827--1044& 1493 1240--176\\\hline
{\tt Approx~10}& 789..9 718--884& 1129.7 980--1292
\\\hline
{\tt Approx~1000}& 704.7 651--759 & 974.1  892--1066
\\\hline
{\tt Exact}& 708 655--769&  698.8 867--1101 \\\hline

\end{tabular}

\begin{tabular}{|l||c|c|}
\hline
 &   $99\%$ &   $100\%$\\\hline\hline
{\tt RW}&$10^6$ $10^5--10^7$ & $10^7$ $10^6 -- 10^7$\\\hline
{\tt Uniform}   &3405.7 2760--4186 & 11049 6563--22480\\\hline
{\tt Approx~10}&1998.8 1707--2282& 11962 3336--62763\\\hline
{\tt Approx~1000}&1625.3 1441--1841& 3410.8 2265--7144\\\hline
{\tt Exact} & 1610.5
1402--1843 & 3268.8 2369--4738\\\hline
\end{tabular}
\begin{tabular}{l}
{\sf Ttp8},   3201 states,\\  6765 transitions
\end{tabular}
\bigskip

\begin{tabular}{|l||c|c|c|c|}
\hline
 r=0&   $90\%$ &   $95\%$ &   $99\%$  \\\hline\hline
{\tt RW}&  505 248--850& 1073 460 -- 1777& 5702.6 1395--16035
 \\\hline
{\tt Uniform}  & 67.6 47--92& 106.6 73--160& 226.6
115-471 \\\hline
{\tt Approx~10}& 53.05 40--73& 73.3 55--106&134.1 90--226
\\\hline
{\tt Approx~1000}& 49.1 39--64& 67.6 51.95& 115.1 79--199
\\\hline
{\tt Exact}& 49.7 39--72 &66.8 51--95 &114.4 76--168  \\\hline

\end{tabular} 
\begin{tabular}{|l||c|}
\hline
 &   $100\%$\\\hline\hline
{\tt RW}&21750 2579 -- 119811\\\hline
{\tt Uniform}   &372.3 165--809\\\hline
{\tt Approx~10}& 193.3 97--439\\\hline
{\tt Approx~1000}&158.1 85--300\\\hline
{\tt Exact} &157.4 91--270\\\hline
\end{tabular}
\smallskip
{\sf Prodcons10},   286 states,  600 transitions



}
\caption{Comparative results (2) for number of generated tests}\label{fig:res2}
\end{figure}

\begin{table}[!b]
\begin{center}
\begin{tabular}{|l|c|c|c|c|}
\hline
Name & States & Transitions & Eccentricity & Nb. of paths\\
\hline
{\sf Barber1}	&15	&18	&5	&74\\
{\sf Berkeley3}	&1376	&3974	&51	&$1,33\ 10^{39}$\\
{\sf Consistency3}	&806	&1206	&600	&$5,63\ 10^{153}$\\
{\sf Csm1}	&24	&57	&8	&$934000$\\
{\sf Dekker1}	&86	&178	&17	&$8,80\ 10^{11}$\\
{\sf Dragon3}	&103	&696	&50	&$2,34\ 10^{93}$\\
{\sf Fms1}	&120	&582	&14	&$1,41\ 10^{20}$\\
{\sf Illinois3}	&103	&307	&100	&$2,23\ 10^{90}$\\
{\sf Kanban1}	&160	&1151	&14	&$3,31\ 10^{20}$\\
{\sf Lift3}	&499	&587	&302	&$7,24\ 10^{59}$\\
{\sf Moesi2}	&22	&43	&11	&$3,84\ 10^{8}$\\
{\sf Prodcons10}	&286	&660	&20	&$3,51\ 10^{7}$\\
{\sf Ttp8}	&3201	&6765	&32	&$4,30\ 10^{7}$\\\hline
\end{tabular}\smallskip

\end{center}
\caption{Graphs used for benchmarking.}\label{tab:protocols}
\end{table}

This section is dedicated to an experimental evaluation of the proposed
approxi\-mation-based approach. In Section~\ref{sec:benchmark} the set of used
automata is described. Section~\ref{sec:xpprotocol} explains the
experimental protocol. Finally, the obtained experimental results are provided
in~Section~\ref{sec:xpresult}, both for the quality of the approach and for
computation time.

\subsection{Benchmark}\label{sec:benchmark}

Experiments have been done on several automata modeling communication
protocols designed for the FAST tool~\cite{DBLP:conf/cav/BardinLP06}
available\footnote{\url{http://www.lsv.fr/Software/fast/examples/examples.tgz}}
online as a library of parametric counter automata (the parameter can be,
for instance, the number of communicating processes). For several examples
and parameters, the counter automaton has been faltered into a classical
finite automaton. The list is given in Table~\ref{tab:protocols}: first
column contains the name of the protocol with the value of the parameter.
The second and the third columns respectively report on the number of states
of the automaton and the number of transitions. The fourth column provides
the eccentricity\footnote{Eccentricity is an important parameter since it is
the minimal length required for paths to have a chance to visit each state.}
of the automaton, that is the maximal distance of an edge to the initial
states. Finally, the last column gives the approximate number of successful
paths in the automaton of length less than or equal to twice the
eccentricity. Note that in these graphs all states are final.

\subsection{Experimental Protocol}\label{sec:xpprotocol}

For each protocol, we have measured the number of tests/generated paths
required to cover either $50\%$, or $90\%$, or $95\%$, or $99\%$, or $100\%$
of the states. Several values close to $100\%$ have been chosen since many
biased approaches have been introduced to handle rare events, and many
methods will efficiently cover $50\%$ or $70\%$ of the graph. It is harder
to cover the remaining last states. We have compared 5 different approaches.
First, the {\tt RW} Approach consists in performing isotropic random walks
in the automaton: once in a state, the next one is picked up uniformly among
its neighbours. The path ends either when it reaches a dead-end state, or
when its length is twice the eccentricity. The second approach, denoted {\tt
Uniform}, is the one introduced in~\cite{DBLP:journals/sttt/DeniseGGLOP12}:
paths of length bounded by twice the eccentricity are uniformly generated.
The approach denoted {\tt Exact} is the biased approach proposed
in~\cite{DBLP:journals/sttt/DeniseGGLOP12}, where the linear system is
exactly computed. The {\tt Approx~10} and {\tt Approx~1000} approaches
are the ones proposed in this article: for $10$ [resp. $1000$] the
$\alpha_{i,j}$'s are approximated using $10n$ [res. $1000n$] randomly
generated paths, where $n$ is the number of states.

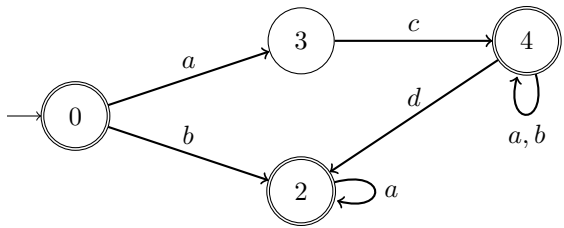
\begin{figure}[!t]
\begin{center}
\begin{tikzpicture}
\node[draw, state, accepting, initial, initial text=] (1) at (0,0) {$0$};
\node[draw, state] (3) at (3,1) {$3$};
\node[draw, state, accepting] (4) at (6,1) {$4$};
\node[draw, state, accepting] (2) at (3,-1) {$2$};

\path[->,thick] (1) edge[above] node {$b$}  (2);
\path[->,thick] (1) edge[above] node {$a$}  (3);
\path[->,thick] (3) edge[above] node {$c$}  (4);
\path[->,thick] (4) edge[above] node {$d$}  (2);

\path[->,thick] (2) edge[loop right] node {$a$}  ();
\path[->,thick] (4) edge[loop below] node {$a,b$}  ();

\end{tikzpicture}
\end{center}
\caption{Illustrating example with different optimal solutions.}\label{fig:exproba}
\end{figure}

Note that comparing the distribution $\pi$ given by the exact approach and
the approximation-based approaches is not easy. Indeed, a linear programming
system may have different optimal solutions. Let us consider for instance the
example depicted in Fig.~\ref{fig:exproba}. The set of successful paths
visiting $3$ is the same as the set of successfully paths visiting $4$.
Therefore, in any optimal solutions of the linear programming system given
$\pi_3=x$ and $\pi_4=y$, one can do the following changes: $\pi_3=z$ and
$\pi_4=t$ with $z+t=x+y$, and we also obtain an optimal solution.

\subsection{Qualitative Experimental Results}\label{sec:xpresult}

Since the test generation procedures are randomized, performance is
stochastic. For each example and each coverage proportion, each approach has
been experimented~100 times. For each case, we report on the average number of
tests obtained in order to cover the wanted proportion, but also the minimum
number of tests (the best case), and the maximum number of tests (the worst
case).

Results presented in Figs.~\ref{fig:res1} and~\ref{fig:res2} are obtained with
$r=0$~: there is no a priori guarantee on the precision of the
approximations. Results presented in Fig.~\ref{fig:res3} are obtained with
$r=0$ and with $r=10$ (and in one case with $r=50$). For instance, the
second table in Fig.~\ref{fig:res1} reports on the result for {\sf Dekker1}: in
order to cover $95\%$ of the set of the states, the {\tt RW} approach
requires on average 790 paths. In the best case (of the experiments), it only
requires 159 paths, and in the worst case 2062 paths have been generated. For
the same coverage, the {\tt Uniform} approach requires 35.45 paths in
average. The {\tt Exact} approach only requires 15.5 paths in average.

Relatively to the other approaches, the performance of {\tt RW} deeply
depends on the topology of the automaton. For instance, for {\sf Prodcons10}
or {\sf Ttp8} or {\sf Moesi2}, {\tt RW} is ugly, and requires much more tests
to (partially) cover the set of states. For {\sf Fms1}, {\tt RW} is as efficient
as {\tt Exact}. For some automata, there is no result for {\tt RW}: after
 hours of computation, the approach was not able to cover 90\% of the set
of states. In these cases, some states occur with a so low probability on
random walks, that in practice it is not possible to generate a path
visiting them.

One can see that for {\sf Barber1}, {\sf Dekker1}, {\sf Fms1}, {\sf Moesi2},
{\sf Kanban1},  {\sf Ttp8} and {\sf Prodcons10}, all biased approaches are
better (cover the set of states with less paths) than the uniform one. Moreover, the
{\tt Exact} approach is better than the approximate ones, but not
significantly with the {\tt Approx~1000}. Consider for instance {\sf Fms1}:
the {\tt Uniform} approach requires on average 87.6 paths to cover 99\% of
the states. With {\tt Approx~10} this number falls to 48.1, and it falls to 40.3
with {\tt Approx~1000}. The {\tt Exact} approach requires 29.8 paths
on average.


The results for {\sf Lift3} are similar but the {\tt Approx~1000} is not so
close to the {\tt Exact} approach. For {\sf Berkeley3}, {\sf Illinois3} and
{\sf Dragon3}, the {\tt Exact} approach is clearly more efficient to cover
the set of states. It is similar for {\sf Consistency3}, but only for the 100\%
coverage criterion. A significant case is {\sf Illinois3}: the {\tt
Exact} approach requires on average a unique path to cover all states, while
the {\tt Approx 1000} approach requires 47 paths. For all these examples there is a
huge number of paths, and many states $j$ are visited with a very low
probability by a path: the corresponding $\alpha_{i,j}$'s are set to zero
since $r=0$, thus providing a very bad approximation. For instance, for {\sf
Illinois3}, 84 states over the 103 states are not visited by any random
paths. We run the experiment with  {\tt Approx 1000} and $r=10$. The obtained results are
presented in Fig.~\ref{fig:res3}: these results are much better and close to the
ones of the {\tt Exact} approach.

In conclusion, for the quality of the coverage, running {\tt Approx~10}
with $r=10$ seems to be an efficient solution.


\begin{figure}[!p]
{\small

\begin{tabular}{|l||c|c|c|c|c|}
\hline
r=0 &  $90\%$ &   $95\%$ &   $99\%$ &   $100\%$ \\\hline\hline
{\tt Uniform}   & 176 153--240 & 315 258--375& 630
486--834& 1203 837--2376\\\hline
{\tt Approx~10}& 166 142--212& 301 253--377& 648
431--884&2127 797--5611\\\hline
{\tt Approx 1000}& 171 259--381& 307 259--381& 673
485--1070& 1973 807 -- 4485\\\hline
{\tt Exact}& 166 143 -- 203&  294 256--346& 589
440--816& 1084 689--1783\\\hline\hline
{\tt Approx~10} $(r=10)$&168 140--231 & 297 256--376 & 626
479--780 & 1474.9 730 -- 4749 \\\hline
{\tt Approx~1000 }$(r=10)$&168 133--211 & 300 258--359 & 624
489--828 & 1390 765 -- 4042 \\\hline
{\tt Approx~10 }$(r=50)$&168 141--208 & 300 250--373 & 610
480--848 & 1292 710 -- 2715 \\\hline
\end{tabular}
\smallskip
{\sf Consistency3},  806 states, 1206 transitions

\bigskip
\begin{tabular}{|l||c|c|c|c|c|}
\hline
r=0 & $90\%$ &   $95\%$ &   $99\%$ &   $100\%$ \\\hline\hline
{\tt Uniform}  & 10.56 4--30& 23 6--100& 64.7 10--400& 85.3 10--400\\\hline
{\tt Approx~10}& 5.4 3--14& 9 4--46& 24.7 8--107& 34.9 9--137
\\\hline
{\tt Approx 1000}& 4.4 3--7& 6.6 4--23& 13.9 6-37& 18.9 7-51
\\\hline
{\tt Exact} & 3.9 3--6& 5.5 4-9& 9.5 6--20&13.1 6-24 \\\hline\hline
{\tt Approx~10} $(r=10)$& 3.8 3--4 & 5.4 4--7 & 8.9 6--15 &12.9 8--23
\\\hline
{\tt Approx 1000} $(r=10)$& 3.7 3--5 & 5.3 4--7 & 9.0 7--14 &
12.4 7-20
\\\hline

\end{tabular}

Lift3,  499 states, 587 transitions
\bigskip

\begin{tabular}{|l||c|c|c|c|c|}
\hline
r=0 &  $90\%$ &   $95\%$ &   $99\%$ &   $100\%$ \\\hline\hline
{\tt Uniform}  & 22.5 7--101& 35.8 12--135& 69.6 24--267&112.1
34--317 \\\hline
{\tt Approx~10}& 18.3 9--47 & 29.1 12--76 &63.2 15--231 &
98.5 34--267
\\\hline
{\tt Approx 1000}& 14.7 6--29& 25.2 11--55& 56.7 13--184 & 87.2
13--319
\\\hline
{\tt Exact} & 10.8 6--17& 15 7--25& 22.8 10--41& 29.9 10--57\\\hline\hline

{\tt Approx~10} ($r=10$)& 10.6 5--17&14.6 6--29 & 23.5 12--47 &
30.7 16--79 
\\\hline
{\tt Approx 1000} ($r=10$)& 10.1 6--18& 14.3 8--26 & 22.6 14--42
& 29.5 14--79
\\\hline
\end{tabular}
\begin{tabular}{l}
{\sf Dragon3},   103 states,  696 transitions
\end{tabular}

\bigskip
\begin{tabular}{|l||c|c|c|c|c|}
\hline
r=0 &  $90\%$ &   $95\%$ &   $99\%$ &   $100\%$ \\\hline\hline
{\tt Uniform}   & 337 235--448& 594 400-863& 1551 993--2262&
4470 2316--9974\\\hline
{\tt Approx~10} & 222 167--286& 384.8 258-536 & 1014.6
655-1382& 2886.6 1504--5251 \\\hline
{\tt Approx 1000} & 189 134--247& 323 232--498& 854
467--1302& 2116 1108--2886\\\hline
{\tt Exact} & 103 60--185 &137.0 84--210& 199.6 109--364&
224.1 128--510 \\\hline\hline
{\tt Approx~10 (r=10)} & 102 66-149 & 138 80-280 & 
207.0 107--399 & 233.6 140--412\\\hline
{\tt Approx 1000 (r=10)} & 102 67--202 & 137 86-256 &
206 112--423 & 231 122--508\\\hline

\end{tabular}

{\sf Berkley3}, 1376 states, 3974 transitions
\bigskip

\begin{tabular}{|l||c|c|c|c|c|}
\hline
r=0 &  $90\%$ &   $95\%$ &   $99\%$ &   $100\%$ \\\hline\hline
{\tt Uniform} & 8.8 1--46 & 16 1-73& 40.1 1--209  & 68.43 1--338\\\hline
{\tt Approx~10}& 7.5 1-27& 12.1 1--46& 30.1 1--150&56.9 1--314
\\\hline
{\tt Approx 1000}& 6.2 1--38 & 11.25 1-61 & 29.1 1-210& 48.3 1--234
\\\hline
{\tt Exact} & 1.0 1-2 &1.0 1-2 & 1.0 1-2& 1.1 1-2\\\hline\hline
{\sf Approx 10 } $(r=10)$ & 1.0 1--2& 1.0 1--2& 1.1 1--2 & 1.2 1--3\\\hline
{\sf Approx 1000 } $(r=10)$ & 1.0 1--2& 1.0 1--2& 1.1 1--2 & 1.2 1--2\\\hline
\end{tabular}
\begin{tabular}{l}
{\sf Illinois3},   1524 states,  307 transitions
\end{tabular}}

\caption{Comparative results (3)}\label{fig:res3}
\end{figure}

\subsection{Computation Time}

\begin{table}[!t]
\begin{center}
\begin{tabular}{|c||c|c|c|c|c|}
\hline
(seconds)& {\sf Berkeley3}  & {\sf Consistency3} &{\sf Dragon3} &{\sf Lift3} & {\sf Illinois3}\\\hline\hline
{\tt Exact}             & 26401 & 40964& 29& 4337& 18\\\hline
{\tt Approx 10} ($r=0$)   & 1   & 58   & 1 & 8   & 0.4\\\hline
{\tt Approx 1000} ($r=0$) & 186 & 5794 & 21& 813 & 39\\\hline
{\tt Approx 10} ($r=10$)  & 16  & 110  & 1& 12   & 1\\\hline
{\tt Approx 1000} ($r=10$)& 208 & 5890 & 25& 862 & 41\\\hline
{\tt Approx 10} ($r=50$)& --  & 124  & --& --  & - \\\hline
\end{tabular}
\end{center}
\caption{Time to compute the linear programming system.}\label{tab:comp}
\end{table}

Let us note first that for all approaches, generating paths is done
practically in a very efficient way. As mentioned before, the bottleneck step is the computation
of the linear programming system. In Table~\ref{tab:comp}, the time (in
seconds) used to compute the linear programming system is given for the
protocols {\sf Berkeley3}, {\sf Consistency3}, {\sf Dragon3}, {\sf Lift3} and
{\sf Illinois3}. The results are similar for the other protocols. For {\sf
Illinois3}, using {\tt Approx~1000} is less efficient than using the {\tt
Exact} approach. The reason is that the automaton is quite small. However,
for other cases, using the approximation-based approaches is faster. And it
is significantly faster for large automata. For instance, for {\sf
Consistency3}, while the {\sf Exact} approach requires more than 11~hours, and only
about 90 minutes are needed for {\tt Approx~1000} (with $r=0$).

In all cases, the best compromise seems to use {\tt Approx~10} with
$r=10$: the computation time is strongly better, and the quality of the
biased approach is similar to the {\tt Exact} approach, except for {\sf
Consistency}. For this protocol, we run the   {\tt Approx~10} with
$r=50$ and we obtain better results, closer to the {\tt Exact} approach,
with a very short computation time (about 2 minutes, in comparison to 11 hours for the {\tt Exact} approach).

\subsection{Experiments on Large Graphs}

We have experimented the approaches on a model of the {\sf
Centralserver2} protocol, which has 2523 states and 18350 transitions,
an eccentricity of 63, and about $8,04\ 10^{113}$ successful paths of
length less or equal to 126.  By computing the first
$\alpha_{i,j}$'s, we estimate that the computation time of the linear
programming system with {\tt Exact} will require about 200~days. The
linear programming system with {\sf Approx~10} and {\sf Approx~1000}
($r=10$) has been computed respectively in 81 seconds and in 24 minutes.
The obtained qualitative results compared to {\tt Uniform} are given in
Table~\ref{res:cs2}.

\begin{small}
\begin{table}
\begin{tabular}{|l||c|c|c|c|c|}
\hline
 &  $90\%$ &   $95\%$ &   $99\%$ &   $100\%$ \\\hline\hline
{\tt Uniform}  & 680 573--786&  1198 947--1492&  2942 2358--3612&  9413 5487--19533\\\hline
{\tt Approx~10} ($r=10$) & 316 287--349&  476 437--529&  878 775--1037&  2065 1223--4337\\
\hline{\tt Approx~1000} ($r=10$) & 313 281--345&  467 415--515&  864 729--1037&  1926 1252--3514\\
\hline
\end{tabular}
\smallskip
\caption{Results for {\sf  Centralserver2}.}\label{res:cs2}
\end{table}
\end{small}

We also used the algorithm proposed in~\cite{DBLP:conf/stacs/CarayolN12} to randomly generate two trim
automata with respectively 5659~states (with 17007~transitions) and
11251~states (with 33753~transitions). The approximated linear
programming system obtained by the {\t Approx~10} and {\tt
Approx~1000} approaches (with $r=10$) has been computed in
respectively 5.5s and 613s for the first graph, and in respectively
26.3s and 1162s for the second graph.

\section{Conclusion}

In this paper we proposed an approximation-based approach for the random
biased exploration of large models. It has been experimented on several
examples: in practice the approximation is not too coarse, and the quality of
the generated test suites to cover the states of the model is excellent
compared to the exact approach and to the other random approaches. For
computation time, using approximation is significantly better since the approach can be used on graphs with more than 10000~states.
In the future we plan to investigate recent advances in optimization in order
to improve the computation time.

\bibliographystyle{plain}
\bibliography{biblio}

\newpage
\end{document}